# High-yield and high-angular-flux neutron generation from deuterons accelerated by laser-driven collisionless shock

C.-K. Huang[a)], D.P. Broughton, S. Palaniyappan, A. Junghans, M. Iliev,
S.H. Batha, R.E. Reinovsky, A. Favalli[b)]

*Los Alamos National Laboratory, Los Alamos, NM, 87545, USA*

## Abstract

A bright collimated neutron source is an essential tool for global security missions and fundamental scientific research. In this paper, we study a compact high-yield and high-angular-flux neutron source utilizing the break-up reaction of laser-driven deuterons in a $^9$Be converter. The neutron generation scaling from such a reaction is used to guide the choice and optimization of the acceleration process for the bulk ions in a low density $CD_2$ foam. In particular, the collisionless shock acceleration mechanism is exploited with proper choice in the laser and target parameter space to accelerate these ions towards energies above the temperature of the distribution. Particle-In-Cell and Monte Carlo simulations are coupled to investigate this concept and possible adverse effects, as well as the contribution from the surface ions accelerated and the optimal converter design. The simulation results indicated that our design can be a practical approach to increase both the neutron yield and forward flux of laser-driven neutron sources, reaching peak angular neutron flux $>10^{11}$ neutron/sr and yield $>10^{11}$ neutron/pulse with present-day kJ-class high-power lasers. Such developments will advance fundamental neutron science, high precision radiography and other global security applications with the laser-driven sources.

a-b) Authors to whom correspondence should be addressed.  Electronic mails: a) huangck@lanl.gov and b) afavalli@lanl.gov.

The development of high intensity neutron sources is opening a wide range of opportunities, such as characterizing static objects and diagnosing dynamic experiments using neutron radiography, detection of special nuclear materials for global security applications, and other fundamental and applied applications [1-5]. In recent years, intense laser-driven neutron sources have attracted significant interest thanks to the advance of ion acceleration by short-pulse high-intensity lasers [6-14]. For example, neutrons with high angular flux ($10^{10}$ neutrons/sr) in a short-duration (~ns) generated by laser-accelerated ions have demonstrated for the first time the active interrogation of nuclear material in a single laser-driven neutron shot [5, 10].

However, many applications for basic science and global security require the development of high energy and collimated neutrons with a higher neutron flux (e.g., greater than $10^{11}$ neutrons per sr per shot [15]), to provide a sufficiently high penetrability in shielding materials. In this paper we introduce the design considerations that enable such neutron sources using the deuteron breakup neutron production and the collisionless shock acceleration (CSA) mechanism. Results of the simulation study and optimization are presented and discussed, where the VPIC code [14] was used for modeling the ion acceleration and coupled with the Monte Carlo N-Particle (MCNP) version 6.2 [16] for the simulation of neutron production and transport.

Designing a laser ion acceleration process with good efficiency, together with a suitable neutron-producing reaction and optimized converter design, is promising for achieving a collimated, high energy and high intensity neutron pulse [17, 18]. The deuteron breakup reaction is favored over D-D fusion reaction for this purpose due to its larger cross-section and the highly directional neutron flux generated [19, 20]. Specifically, the thick target neutron yield per incident deuteron of kinetic energy $E_d$ and the forward neutron flux per steradian scales as $E_d^2$ and $E_d^{2.5}$, respectively (see Supplementary Material A for MCNP simulation verification) [21-24]. One challenge with the currently realized laser-driven deuteron acceleration mechanisms [25, 26], such as Target Normal Sheath Acceleration (TNSA) and Relativistic Transparency (RT), is the production of a broad exponential energy spectrum with a low bulk deuteron energy that is less efficient for the generation of an energetic and collimated neutron beam. In particular, the record-setting neutron experiments in the RT regime generated deuteron beams with an average energy of ~4 MeV with 100% energy spread [12]. Similarly, for TNSA-based neutron sources, while both neutron yield and forward flux increase with the ion temperature $T$ as $\sim T$ and $\sim T^{2.5}$

(Supplementary Material A), much of the laser energy is converted into the low energy ions below the temperature of the distribution. Besides the consideration of the distribution of the accelerated deuterons, achieving high yield from the acceleration process is another key optimization strategy. Furthermore, neutron source size is an important factor for practical applications and laser-driven neutron source using a conventional pitcher-catcher setup is typically on the order of mm [5], partly because the catcher needs to be placed at a sufficient distance from the pitcher to avoid interruption of the acceleration process at the back side of the target.

In light of above considerations, an alternative way to exploit the deuteron breakup reaction is to combine the Hole-Boring Radiation Pressure Acceleration (HB-RPA) or CSA ions in a near-critical target (e.g., using foam targets or frequency doubled laser) with an attached or minimally separated converter (Fig. 1). This design offers both *high conversion efficiency* and *small source size* as the HB-RPA or CSA mechanism can accelerate a large fraction of the target ions within the laser spot area to energies beyond the temperature of the distribution, simultaneously improving the useful neutron yield and forward flux (see details in Supplementary Material B). As the majority of the acceleration process happens inside the target, a converter separately optimized can be used immediately behind the target for a compact design.

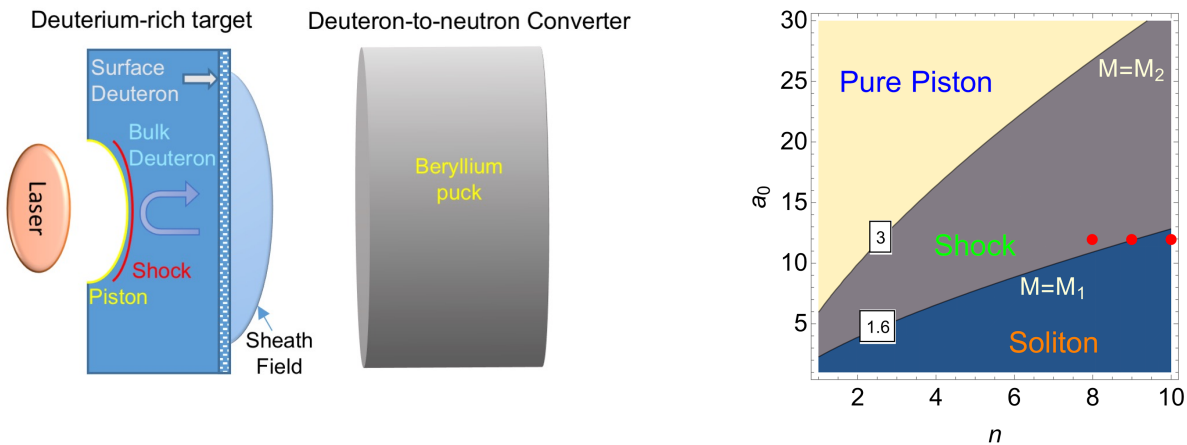

**Figure 1** (*Left*) The schematic of a pitcher-catcher neutron source using ions accelerated in a low-density deuterium-rich target by a laser driven piston or collisionless shock. The $^9$Be converter (not to scale) is optimized based on the accelerated deuterons. (*Right*) The shock regime of ion acoustic wave by a laser-driven piston in a deuteron target is illustrated in the parameter space of electron number density $n = n_e/n_c$ and normalized laser field strength $a_0$, assuming the Haines hot electron temperature scaling [27] and a linearly-polarized laser. The boundaries of the shock regime are denoted by nominal critical Mach numbers $M = M_1 = 1.6$ and $M = M_2 = 3.0$ [28, 29]. A typical target reflectivity $R = 0.4$ is used. Our PIC simulation designs are shown by the red dots.

When a laser with intensity $I$ drives a piston at velocity $v_b = \beta c$ and bores a hole into the target with ion density $n_i$ and mass $m_i$, a collisionless shock can be formed from the steepening of the nonlinear ion acoustic wave when the piston Mach number $M = v_b/u_s$ ($u_s$ and $c$ are the local ion sound speed and speed of the light, respectively) falls between two critical Mach numbers, $M_1 < M < M_2$ [27, 28]. Bulk ions in the target can be reflected by the shock electric field to twice its speed, thus collisionless shock acceleration of deuterons is particularly attractive for our purpose. As $\beta \approx \sqrt{(1+R)\Pi/2}$ (see Supplementary Material C, here $\Pi = I/m_i n_i c^3$ and $R$ is the target reflectivity measured in the lab frame), the piston Mach number $M$ depends on the laser intensity and target density. Therefore, the regime of CSA can be characterized in the parameter space of $(a_0, n)$ as shown in Fig. 3 right, where $a_0 \approx 8.6 \times 10^{-10} \lambda_L[\mu m] I^{1/2}[W/cm^2]$ and $n = n_e/n_c$ are the normalized peak laser electric field and target electron density respectively. For the CSA, high $a_0$ and low $n$ are needed to achieve a faster shock thus higher deuteron energy. However, at high $a_0$ (i.e., closer to the upper boundary in Fig. 1 right), highly nonlinear laser plasma interactions can interrupt the shock reflection of ions. Particularly, relativistic transparency can lead to laser penetration in the piston, while laser filamentation/self-focusing can occur when a wide laser spot size is used in conjunction with high $a_0$ [30]. Therefore, we choose $a_0$ and $n$ near the lower boundary of the shock regime [31] where the reflected ions can still reach reasonably high energies. In order to selectively transfer shock energy to the deuteron ions, ideally other ions in the region the shock traverses should have a lower charge-to-mass ratio, hence high ionization charge states from tightly-focused lasers should be avoided. Another parameter that is implicitly included in the hole-boring velocity and the regimes shown in Fig. 1 right is the laser pulse length, where the assumption is that the ions upstream will have sufficient time to respond and reach the reflection speed. In fact, the inertia of the ions, in conjunction with the actual laser pulse shape, sets a limit of how long a laser pulse needs to be in order to have the piston moving at the expected speed. With these considerations in mind, we conducted a series of PIC simulations (shown as red dots in Fig. 1 right that are discussed below) to identify CSA deuteron acceleration favorable for neutron generation without adverse effects.

In our 2D VPIC simulations, low density deuterated plastic (CD₂) foam targets [33] with $n = 8,9,10$ are chosen for a circularly-polarized driving laser focused to a spot size of $r_0 = 10 \mu m$ and a peak intensity of $I_0 = 3.6 \times 10^{20} W/cm^2$. The laser has a central wavelength of $\lambda = 1 \mu m$, $a_0 = 11.5 \approx 12$, total duration of $\tau = 1$ ps with an intensity profile $I(t) = I_0 \, sin^2 \, (\pi t/\tau)$ for $0 < t < \tau$, and energy of 282J. All targets simulated have 20 micron thickness designed to match the piston travel distance during the laser pulse. The first 18 micron layer at the laser facing side is referred to as the "bulk" where the internal fields provide the dominant acceleration, while the last 2 micron layer is referred to as "surface" where the ions are mostly accelerated by the backside target normal sheath field from the escaped hot electrons. We assume that the Carbon ion is ionized to $C^{4+}$ which is justified by the large ionization energies of the inner shell electrons.

Among the 3 different targets simulated, we found that the $n = 9$ target provides the most stable acceleration and the highest neutron yield. For the $n = 8$ target, laser filamentation is observed, while for the $n = 10$ target the deuteron energy is relatively low and leads to less efficient neutron generation. We further conducted the simulations for the $n = 9$ target using two types of transverse laser intensity profiles, i.e., $I(r) = I_0 e^{-(r/r_0)^m}$ with $m = 2$ (standard Gaussian) or $m = 4$ (super-Gaussian). The former has a less uniform transverse laser intensity profile, hence a pronounced multi-dimensional effect that makes the deuterons more divergent and degrades the deuteron spectral shape (hence the neutron yield and forward flux per deuteron) from the ideal 1D scenario. Fig. 2 summarizes the results from our simulations with $n = 9$ CD₂ target for the bulk and surface deuterons and for the cases with $m = 2,4$. In the bulk deuteron spectrum collected within $\pm 20°$ (green and red curves), a plateau formed by ion reflection within the target and a high energy tail from (both internal and backside) sheath field acceleration can be clearly identified. The exponential tail of the distribution has a temperature of about 20 MeV. The plateau is around 20-60 MeV (or 1-3 times the tail temperature) for the $m = 4$ super-Gaussian laser profile, while it is at about 20-35 MeV for the case with the $m = 2$ Gaussian profile. When collecting all deuterons in the forward direction, the plateau (blue curve for $m = 4$) in the bulk deuteron spectrum is still visible although less pronounced, again due to the multi-dimensional effect in acceleration. Compared to the surface deuteron distribution, the bulk distribution is nearly one order of magnitude higher because of the higher number of ions in the larger target volume. However, this deuteron beam has a larger divergence angle as shown in the energy-angle plot in Fig. 2, while the surface deuterons accelerated are more collimated.

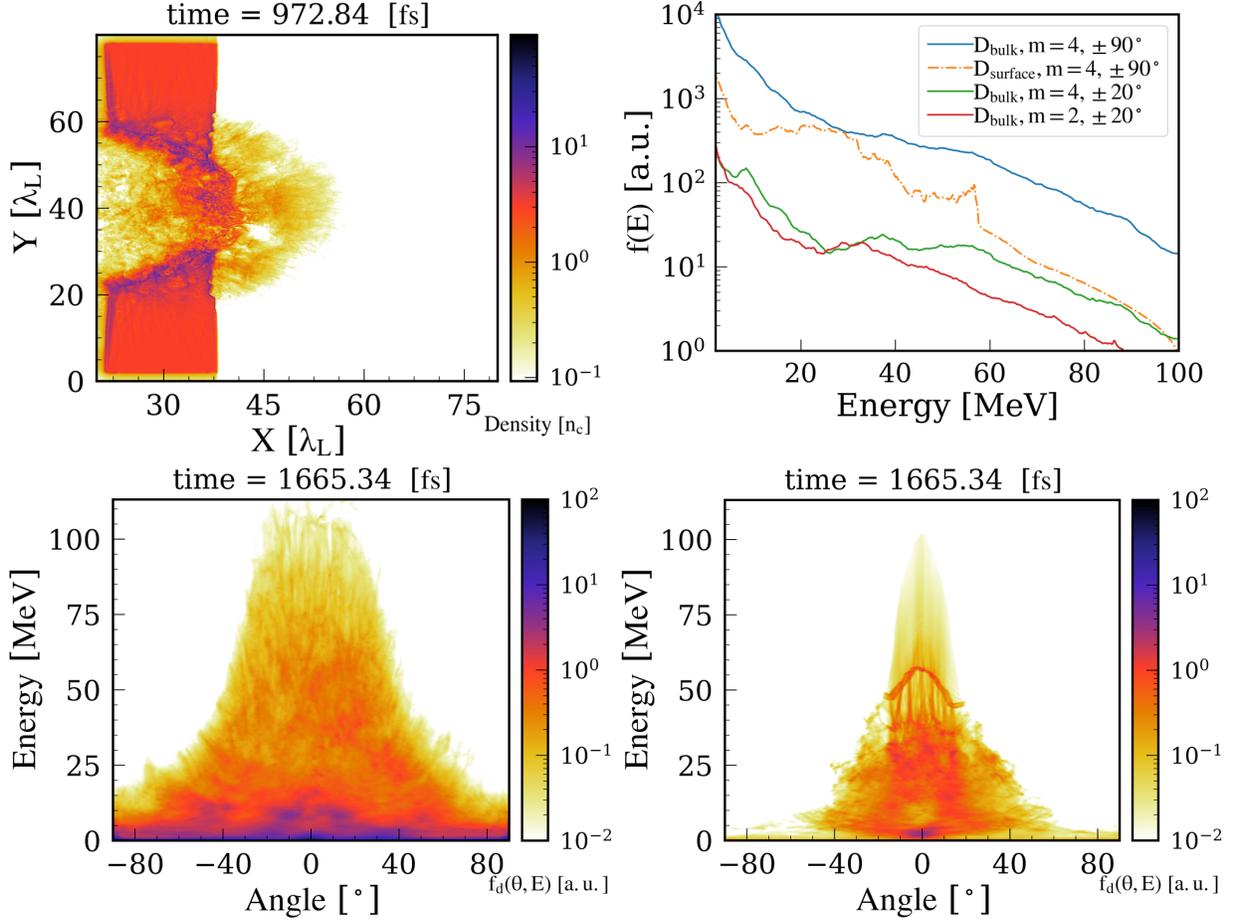

**Figure 2** PIC simulation results for the $n = 9$ CD$_2$ target. Number density of bulk deuterons accelerated from inside the target is shown at time $t = 972.84$fs when the piston breaks through the target (*top left*). The laser impacts the target initially located between $20\mu m < X < 40\mu m$ from the left side. The simulation box is $150\mu m \times 80\mu m$. (*Top right*) The distributions at later time $t = 1665.34$ fs for the bulk deuterons within $\pm 20°$ (red/green curves for $m = 2,4$ respectively) and $\pm 90°$ (blue curve, $m = 4$) are shown for comparison. The orange dashed curve is the surface deuteron distribution within $\pm 90°$. (*Bottom*) The energy-angle distributions of the bulk (*left*) and surface (*right*) deuterons at $t = 1665.34$ fs.

Depending on the application, a converter may be closely positioned next to the $CD_2$ foam to utilize the bulk deuterons, e.g., for the demonstration of the bulk acceleration process (with the TNSA field suppressed) or for the trade-off with a more compact source. Here, the resulting energy and angular distributions of both the accelerated bulk and surface deuterons from PIC simulations at $t = 1665.34$ fs when they have moved away from the target were used as the source input into the MCNP simulations. Specifically, each PIC deuteron distribution was divided into 16 angular bins in the forward direction (from 0° to |90°|) and these energy and angular distributions were used in MCNP. For each case $^9$Be converters were simulated as cylinders where the thickness was defined by range, including longitudinal straggling, for deuterons with a range of energies (10, 20, 30, 40, 50, 60, 70, 80, 90, 100, 110, 120 MeV), and the radii was simulated both as being equal to the thickness and based on the Serber angle calculation (discussed in Supplementary Material A).

Results summarized in Table 1 demonstrate the directionality of all sources and also the high neutron yields using the optimized deuteron acceleration schemes (experimental use of a filter between the deuterated target and the $^9$Be converter could be required and the effects of such a filter are evaluated in Supplementary Material D). For all cases the yield [n/sr] is at least three times higher within the 5° cone relative to all forward directions. The largest fraction of collimated forward neutron flux is seen for surface accelerated deuterons (n=10), $D_{surface,n=10}$, where the yield [n/sr] within 5° cone is over fivefold that in all forward directions, a consequence of the high collimation and high energy of the initial deuteron distribution driving neutron production. The highest neutron yields are seen using the bulk accelerated deuterons with the target of n=9, and the flat super-Gaussian laser profile (last row) further increases the yield by a factor >2 relative to the standard bulk deuteron acceleration case (first row), for a maximum yield of ~$7.3 \times 10^{11}$ n/pulse (within $2\pi$). These high neutron yields are a result of the deuteron yield being an order of magnitude greater for bulk relative to surface acceleration because the bulk volume is nearly an order of magnitude (~9×) higher than the volume which the surface acceleration acts upon. Results for $D_{surface,n=10}$ are shown in Fig. 3 alongside results using the bulk deuterons produced using a super-Gaussian beam profile, $D_{bulk,SG,n=9}$, which also has the highest ($2\pi$) forward directed neutron yield. All of the neutron spectra are relatively high energy with plateaus in the range of one to tens of MeV, supporting applications in need of high penetrability due to thick shielding. As the production of neutrons in the forward direction is most probable for high energy deuterons, there is a strong negative correlation between the most probable fast neutron energy (Fig. 3 left) and the degree to which the neutron distribution is forward peaking (Fig. 3 right).

The angular neutron flux within the 5° forward cone is greater for all cases using the $^9$Be catcher with equal height and radius, while the total forward directed neutron yield is greater using the thinner catcher where the radius is based on the Serber angle. This is likely a consequence of the primarily forward directed high energy deuterons contributing to the forward flux within 5°, while lower energy deuterons incident on the converter at higher angles contributing mainly to the $2\pi$ forward directed neutron emissions. As results were reported only for the ideal converter designs, yields for the full sets of $D_{bulk,SG,n=9}$ and $D_{surface,n=10}$ converters are shown in Fig. 4 to demonstrate the importance of converter optimization.

| Deuteron acceleration scheme | $^9$Be converter design | Collimated within 5° cone | | Forward directed ($2\pi$) | |
|---|---|---|---|---|---|
| | | Maximum yield $^9$Be thickness [cm] | Angular neutron flux [n/sr] | Maximum yield $^9$Be thickness [cm] | Neutron yield [n/pulse] |
| $D_{bulk,n=9}$ | Serber | 1.597 | $1.636\times10^{11}$ | 2.529 | $3.399\times10^{11}$ |
| | H=R | 1.597 | $1.752\times10^{11}$ | 1.597 | $3.273\times10^{11}$ |
| $D_{bulk,n=10}$ | Serber | 0.8692 | $9.811\times10^{10}$ | 1.209 | $1.972\times10^{11}$ |
| | H=R | 0.8692 | $1.016\times10^{11}$ | 0.8692 | $1.927\times10^{11}$ |
| $D_{surface,n=9}$ | Serber | 1.209 | $9.811\times10^{10}$ | 2.042 | $1.248\times10^{11}$ |
| | H=R | 1.209 | $1.001\times10^{11}$ | 1.597 | $1.199\times10^{11}$ |
| $D_{surface,n=10}$ | Serber | 1.209 | $7.818\times10^{10}$ | 2.042 | $9.353\times10^{10}$ |
| | H=R | 1.209 | $7.948\times10^{10}$ | 1.597 | $8.989\times10^{10}$ |
| $D_{surface,SG,n=9}$ | Serber | 1.209 | $1.438\times10^{11}$ | 2.529 | $1.801\times10^{11}$ |
| | H=R | 1.209 | $1.474\times10^{11}$ | 1.597 | $1.730\times10^{11}$ |
| $D_{bulk,SG,n=9}$ | Serber | 1.597 | $4.397\times10^{11}$ | 3.059 | $7.254\times10^{11}$ |
| | H=R | 2.042 | $4.679\times10^{11}$ | 2.042 | $7.013\times10^{11}$ |

**Table 1** Summary of simulated maximum neutron yields using bulk deuteron acceleration, $D_{bulk}$, and surface deuteron acceleration, $D_{surface}$, with standard and super-Gaussian (SG) beam profiles for target density $n = 9,10$. The highest neutron yield and angular flux are obtained from the bulk deuterons in $n = 9$ target accelerated with a SG laser beam (last row).

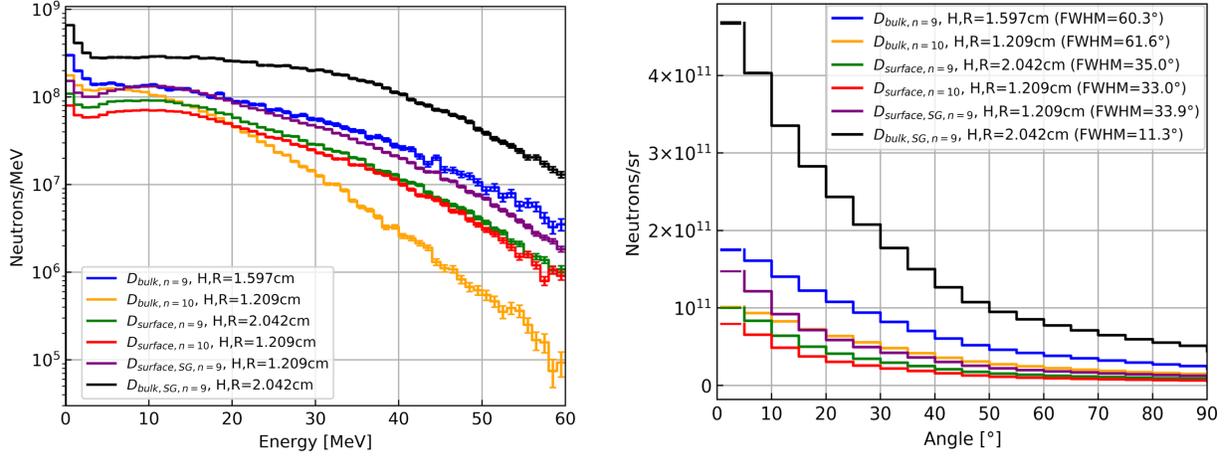

**Figure 3** MCNP results for (*left*) neutron spectrum within 0-5° forward cone, and (*right*) neutron angular distribution (binwise normalization per sr, i.e., using $2\pi(cos(\Theta_{max}) - cos(\Theta_{min}))$ with angular width $(\Theta_{max} - \Theta_{min}) = 5°$ for each bin, FWHM estimated by linear interpolation between data points). Results shown for the $^9$Be converters with equal height and radius having maximum angular neutron flux within a 5° cone. The neutron spectrum is relatively flat in the 5-20 MeV energy range and the SG laser produces the narrowest neutron beam FWHM angular width of 11.3°.

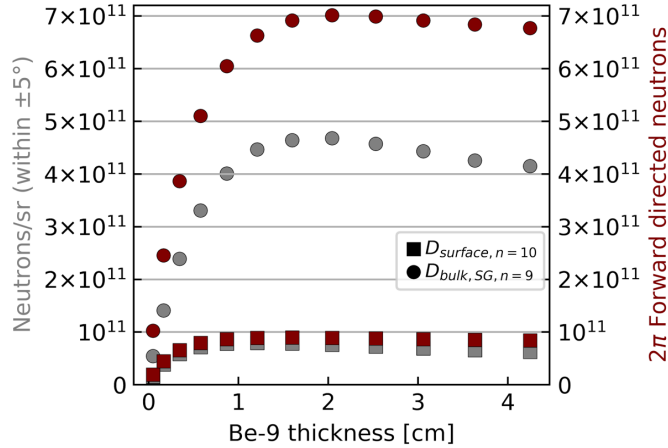

**Figure 4** Angular neutron flux within a 5° cone (*left vertical axis and grey symbols*) and neutron yield in forward direction (within 2π) (*right vertical axis and brown symbols*) are optimized for converter thicknesses (thickness and radius are equal) matched to different deuteron energies. Results using the $D_{surface,n=10}$ and $D_{bulk,\ SG,\ n=9}$ are shown for comparison, the latter produces substantially higher yield and angular flux.

In summary, the neutron generation scaling from the deuteron break-up reaction suggests that choosing and optimizing the ion acceleration process for the bulk ions towards energies above the temperature of the distribution can be a practical approach to increase both the yield and forward flux of laser-driven neutron sources. Furthermore, high laser plasma coupling efficiency of the ion acceleration scheme and a proper design of the neutron converter also play essential roles. In particular, we have demonstrated using VPIC and MCNP simulations the optimization of such a laser-driven neutron source where the bulk deuterons from a $CD_2$ low density foam are mainly accelerated by a collisionless shock inside the target. Despite the non-ideal transverse profile of a focused laser beam leading to multi-dimensional effect in the shock acceleration, the simulation results indicate that ~$1.8\times10^{11}$ n/sr (within 5° cone) and ~$3.4\times10^{11}$ n/pulse can be achieved from the bulk ions using a 282 J laser and an optimized converter. These results are improved to ~$4.7\times10^{11}$ n/sr (within 5° cone) and ~$7.3\times10^{11}$ n/pulse respectively for a laser with a transverse super-Gaussian profile. Furthermore, when the surface deuterons accelerated are taken into account (e.g., with a sufficient gap between the pitcher and catcher), $1.5\times10^{11}$ n/sr (within 5° cone) and $1.8\times10^{11}$ n/pulse can be additionally produced with a potential trade-off in source size. Extensions of this work may explore further optimization of converter design to reduce spot size, which could involve alternative geometries, such as a cone or hemisphere. The design optimizations presented here offer new development opportunities for such neutron sources for experimental applications requiring high-yield and high-forward-flux.

**SUPPLEMENTARY MATERIAL**

See supplementary materials for verification of the neutron generation scaling in MCNP simulations, comparison of deuteron distributions for neutron generation, hole-boring velocity of the laser-driven piston with reflection, and multidimensional effects in deuteron distribution on the normalized neutron yield and forward flux.

**ACKNOWLEDGMENTS**

This work was supported by the U.S. Department of Energy through the Los Alamos National Laboratory (LANL). LANL is operated by Triad National Security, LLC, for the National Nuclear Security Administration of the U.S. DOE (Contract No. 89233218CNA000001). This work was initiated under the LANL LDRD 20180732ER program and continued under OES ADP. C.-K. H. also acknowledges additional support from LANL LDRD project 20140483ER.

## AUTHOR DECLARATIONS

The authors have no conflicts to disclose.

## AIP PUBLISHING DATA SHARING POLICY

The data that support the findings of this study are available from the corresponding authors upon reasonable request.

similar experimental study [32] and the critical Mass numbers in the literature. Higher (lower) $R$ value will increase (decrease) the piston speed and shift the boundaries downwards (upwards). Similarly, hotter (colder) electron temperature will increase (decrease) the ion acoustic wave sound speed and shift the boundaries upwards (downwards).

**Supplementary Material**

## A. Verification of neutron generation scaling with MCNP simulations

For designs involving laser-accelerated deuteron (D) beams, when the high energy deuteron beam bombards a suitable converter material (e.g., $^9$Be), MeV neutrons can be generated mainly via the deuteron breakup reaction. In general, neutron yield increases with laser intensity $I$, as can be understood by the energy scaling of the maximum ion energy $E \sim I^\alpha$ where $\alpha \sim 1/3 - 1$ for various laser ion acceleration mechanisms [25, 26], as well as the correlation between the deuteron energy and the deuteron breakup reaction cross section [19, 20]. Specifically, the thick target neutron yield per incident deuteron of kinetic energy $E_d$ via the deuterium breakup reaction scales as $E_d{}^2$ in a $^9$Be converter [21, 22], as a result of the increased deuteron range and neutron production cross section. In addition, angular divergence of the neutron beam scales as $E_d{}^{-1/2}$, i.e., neutron beam diameter scales as $E_d{}^{1/2}$ (or $E_d{}^{-1/2}$) and source area scales as $E_d{}^1$ (or $E_d{}^{-1}$) for a thick converter designed for the deuteron range (or fixed converter thickness). Overall, the neutron flux density per incident deuteron increases roughly as $E_d$, or even more rapidly depending on the converter design, while the forward neutron flux per steradian scales as $E_d{}^{2.5}$ for a thick converter [22-24].

By assuming a uniform deuteron beam with no divergence and integrating the above scaling relations, it can be estimated that, for a normalized Boltzmann distribution $f(E_d) = e^{-E_d/T}/T$, the total neutron yield and forward flux per deuteron respectively scale as $T$ and $T^{2.5}$. Such an exponential spectrum is a good approximation for many laser-driven ion sources, particularly for the TNSA scheme where the overall ion acceleration efficiency can be relatively high at a few percent.

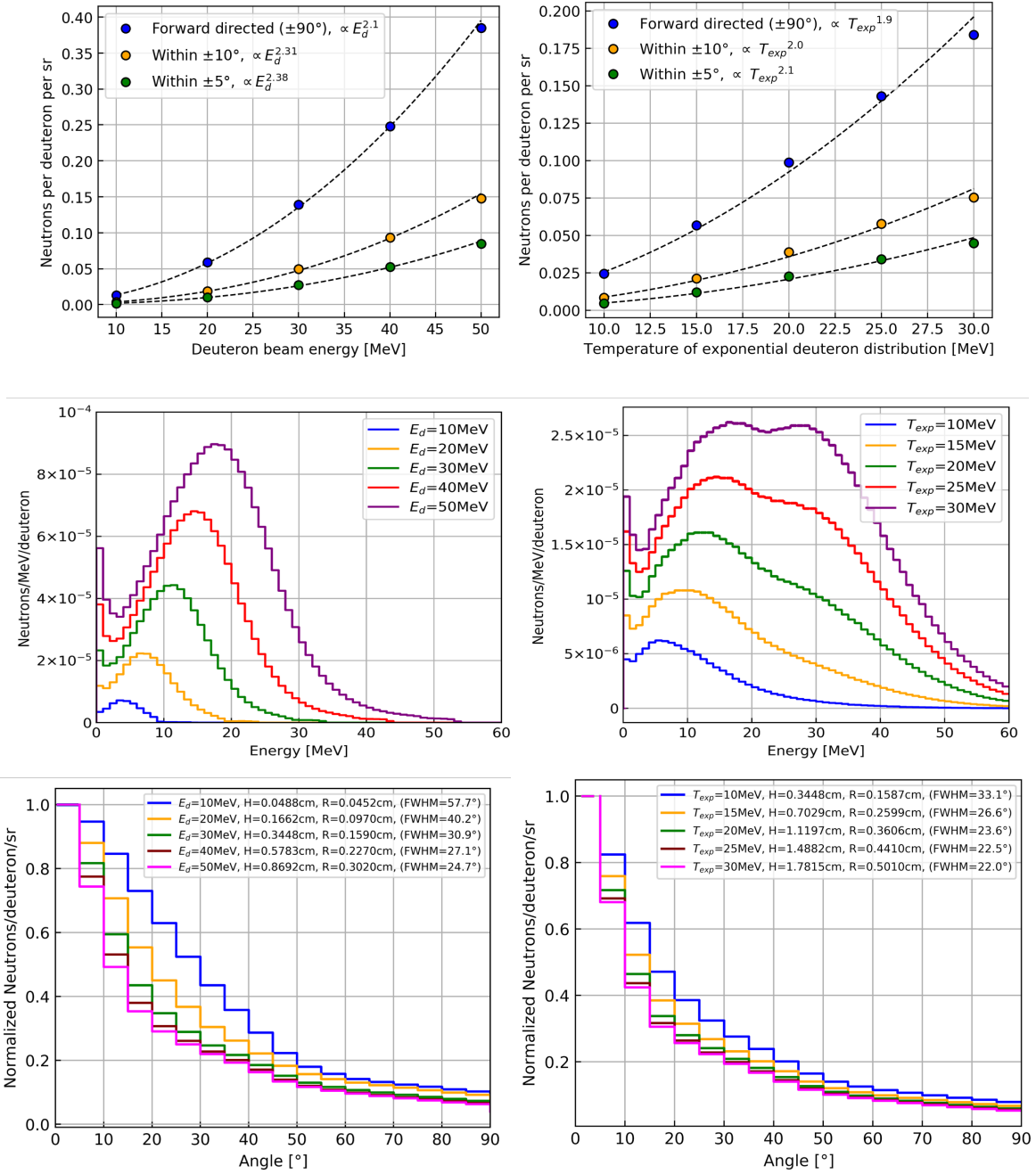

**Figure A1** Verification of (*Top*) the scaling of neutron production and the corresponding (*middle*) neutron energy spectra and (*bottom*) angular distributions (normalized to one to emphasize relative angular distribution, FWHM also shown in legends) through MCNP simulations. Left and right columns are respectively for monoenergetic and exponential deuteron distributions (with various temperature $T$ for the exponential spectrum $f(E_d) = e^{-E_d/T_{exp}}/T_{exp}$). Converter heights and radii in the bottom legend apply to each respective column.

As a verification of MCNP we conducted simulations for a thick converter. The MCNP code is coupled with the evaluated JENDL/DEU-2020 nuclear data for deuteron induced reactions on $^9$Be [34] and a combination of the ENDF70 [35] cross section and be.10t S($\alpha$,$\beta$) thermal molecular scattering data table (both based on ENDFVII.0) [36] to respectively support simulation of neutron production and transport. This recent cross section library was used as it has been shown to enable MCNP simulations predicting neutron yields as a function of both energy and angle consistent with experimental data [34]. Simulations were conducted for both monoenergetic deuteron beams (10, 20, 30, 40, 50 MeV) and exponential distributions of temperatures ($T_{exp}$ =10, 15, 20, 25, 30 MeV) impinging normally on the surface of a $^9$Be converter. As our scaling estimate is most applicable to the neutrons generated on-axis, deuteron beams with no divergence are used in the simulations but neutron angular distributions are tracked. Each converter was cylindrical with thickness corresponding to the initial deuteron range plus the longitudinal straggling distance (for exponential distributions upper energy was chosen to include 95% of the distribution). The radius of each cylinder was determined using this range-based thickness and the Serber stripping angle, $\Delta\theta_{(1/2)} = 1.6(\epsilon_d /E_d)^{0.5}$ where $E_d$ is the kinetic energy of the deuteron and $\epsilon_d$ is its binding energy [23].

The simulation results are consistent with the Serber scaling (Fig. A1 upper left). They also support the derived expression of neutron angular flux for an exponential distribution with temperature dependence proportional to $\sim T_{exp}^{2.5}$ (Fig. A1 upper right), and provide the additional detail that such dependence of the neutron flux is greater for small forward angles ($\sim T_{exp}^{2.1}$) relative to all forward angles ($\sim T_{exp}^{1.9}$). When combined with the fact that the Gamow peak (Supplementary Material B) is at a few times the temperature of this distribution, these are strong motivations to maximize both the energy and number of bulk deuterons to produce the necessary intense pulse of collimated neutrons for applications such as neutron radiography.

In terms of neutron energy, the middle left plot in Fig. A1 demonstrates that the peak shifts towards higher energies, with a peak neutron energy ranging 33% to 38% of the incident deuteron energy, while the peaks are much broader for the exponential deuteron distribution (Fig. A1 middle right). The lower left plot of Fig. A1 clearly shows the angular distribution of these neutrons becomes highly collimated as neutron energy increases, with the FWHM decreasing from 58° at 10MeV to 25° at 50MeV. Variations are smaller for the plots on the bottom right of Fig. A1 using more realistic exponential deuteron distribution because variations in $T_{exp}$ lead to less dramatic shifts in the initial deuteron distribution.

## B. Comparison of distributions and neutron yield/forward-flux for laser accelerated ions

In contrast to the efforts to use target engineering and more delicate processes to produce quasi-monoenergetic or high energy ion beam [37-39], the exponential ion spectrum is quite ubiquitous and robust with solid targets (e.g., solid deuterated plastics or deuterium ice are used for the purpose of neutron generation). One way to enhance the neutron production from the exponential spectrum is to produce hotter deuterons, i.e., by focusing the laser to higher intensity thus exploiting the corresponding laser ion acceleration scaling. However, Fig. A2 shows the relative contribution to the neutron yield and forward flux per steradian for deuterons at different energies (normalized to the temperature $T$) in the exponential distribution. The contribution for the yield (forward flux) peaks at deuteron energy that is 2 (2.5) times the deuteron temperature, a phenomenon similar to the well-known Gamow peak of D-D nuclear fusion in a thermal plasma [40]. Furthermore, in laser-driven ion sources, it is much more efficient to accelerate bulk ions in the solid target to energies around the temperature of the distribution than to the high energy tail of the distribution. Therefore, increasing the number of accelerated deuterons around those energies can be a practical way to optimize the neutron yield and forward flux, requiring modest improvement for the overall laser-to-ion energy conversion efficiency.

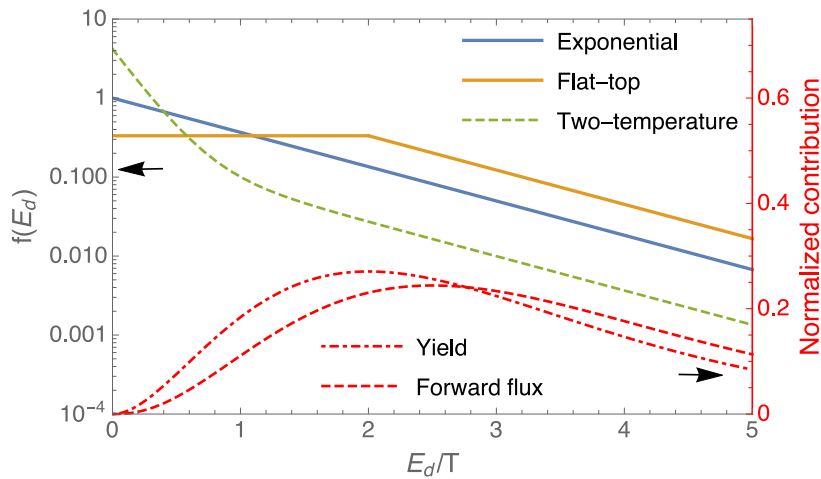

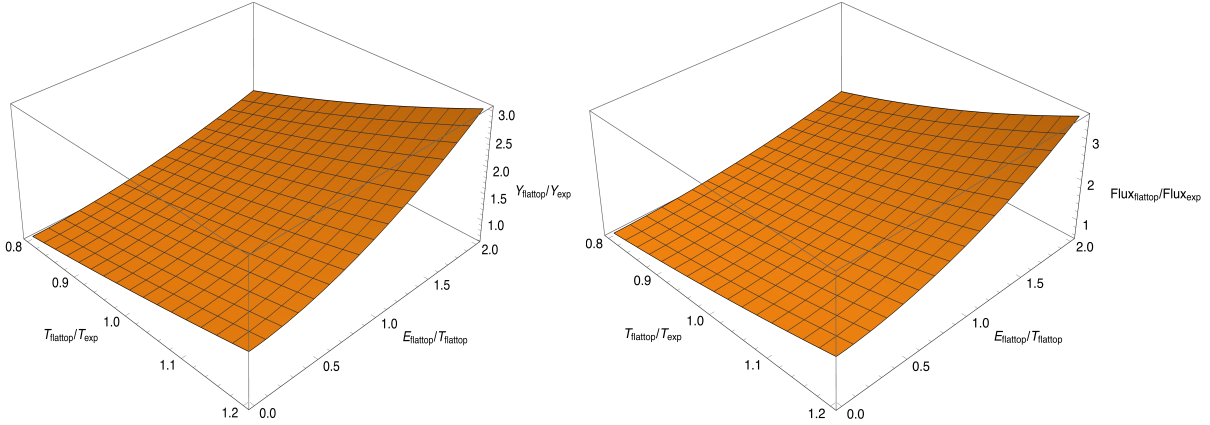

**Figure A2** (*Top*) Three variants of exponential-like (normalized) energy spectra commonly achieved in laser-driven ion sources, i.e., exponential spectra with one temperature, two-temperature and a flat-top, are shown in semi-log scale. The normalized neutron yield (red, dot-dashed) and forward flux (red, dashed) are also shown for the exponential deuteron spectrum. The areas under all curves are normalized to 1. (*Bottom*) Both the neutron yield (*left*) and forward flux (*right*) are enhanced for the flat-top exponential distribution over the one-temperature exponential distribution when the ratios between the tail temperature/flat-top energy and the exponential distribution temperature are increased.

Thus, to achieve high neutron yield and forward flux, we can use the neutron generation scaling to guide the laser target and neutron converter designs. We first consider the energy distribution of the accelerated ions. Two types of distributions (shown in Fig. A2) deviated from the exponential one, i.e., (1) a two-temperature exponential distribution; (2) a flat-top exponential distribution, can be found in realistic laser-driven ion acceleration experiments. Based on the above understandings, additional parameters presented in these distributions can be optimized for the neutron generation. In this work, we focus on the flat-top exponential distribution which has been achieved using low density targets such as nanofoams. This normalized distribution can be parametrized by the energy $E_{flattop}$ (denoting the width of the flat-top) in addition to the temperature $T_{flattop}$ for its tail,

$$f_{flattop}(E_d) = e^{[E_{flattop} - g(E_d)]/T_{flattop}}/(E_{flattop} + T_{flattop}),$$

where $g(E_d) = Max(E_d, E_{flattop})$.

Using the Serber scaling, Fig. A2 also shows the enhancements in neutron yield $Y_{flattop}$ and forward flux $Flux_{flattop}$ for this distribution over those for the exponential distribution of a single temperature $T_{exp}$. It can be seen that with a higher tail temperature or a wider flat-top in the bulk (i.e., larger $E_{flattop}$), both neutron yield $Y_{flattop}$ and forward flux $Flux_{flattop}$ increases beyond the results for the exponential distribution. We note that for results shown at the right corners of the plots, i.e., for $T_{flattop}/T_{exp} = 1.2$ and $E_{flattop}/T_{exp} = 2.0$, both enhancements in the yield and forward flux are higher than 3.0, while the average deuteron energy is about twice as high. Hence, this distribution offers advantage beyond what is indicated by the average energy.

### C. Hole-boring velocity of a laser-driven piston with reflection

The hole-boring velocity $v_b$ given by literature, e.g., Ref [26], does not include reflection coefficient $R$ measured in the lab frame. Here, we derive the hole-boring velocity that includes this coefficient. We denote quantities in the piston frame of velocity $\beta = v_b/c$ with primes. Assuming the incident and reflected lights are plane electromagnetic waves of constant amplitude in vacuum, their time-averaged intensities $I$ and $I_R$ in the lab frame are transformed to the piston frame as $I' = I(1 - \beta)/(1 + \beta)$ and $I_R' = I_R(1 + \beta)/(1 - \beta)$. In the piston frame, the momentum change (per unit surface) in unit time due to the incident and reflected electromagnetic waves is,

$$\Delta p'/\Delta t' = (I' + I_R')/c = [(1 - \beta)/(1 + \beta) + R(1 + \beta)/(1 - \beta)]I/c. \qquad (A1)$$

where we use the reflection coefficient measured in the lab frame $R = I_R/I$. Balancing the momentum change (per unit surface) in unit time from the ion reflection at the piston, $\Delta p'/\Delta t' = 2\gamma^2\beta^2 m_i n_i c^2$, we obtain,

$$\Pi = 2\beta^2/[(1 + R)(1 + \beta^2) + 2(R - 1)\beta]. \qquad (A2)$$

When $\beta \ll 1$, we can drop the last term in the denominator above for typical situations when there is modest reflection.

$$\Pi \approx 2\beta^2/[(1 + R)(1 + \beta^2)], \qquad (A3)$$

which then gives $\beta = \sqrt{(1 + R)\Pi/[2 - (1 + R)\Pi]}$.

Note that we calculate the momentum change of the photon in the piston frame and the reflection coefficient is defined in the lab frame instead for convenience. Furthermore, in the piston frame, the incident and reflected electromagnetic waves have the same frequency $\omega'$, so the frequency transforms to the lab frame differently than the intensity does, i.e., via the relativistic Doppler shift from a moving mirror, $\omega' = \omega\gamma(1 - \beta)$ and $\omega_R = \omega'\gamma(1 - \beta) = \omega(1 - \beta)/(1 + \beta)$. It can be shown that, when $\Pi$ is small, our result gives $\beta \approx \sqrt{(1 + R)\Pi/2}$, in agreement with the result in Ref [26] for $R(\omega') = 1$, i.e., $\beta = \sqrt{\Pi}/(1 + \sqrt{\Pi}) \approx \sqrt{\Pi}$.

### D. Multidimensional effects in deuteron distribution on the normalized neutron yield and forward flux

A curved piston driven by a laser with non-uniform transverse intensity will lead to a divergent deuteron beam deviated from the assumption used in the scaling of neutron production. This can make the spectral plateau feature of the accelerated bulk ions less pronounced and also reduce the number of such ions at high energy. Here, the normalized (per deuteron) neutron forward fluxes are compared for (1) an ideal zero-divergence deuteron beam with an exponential distribution and deuteron beams from bulk ions driven by lasers with (2) the standard Gaussian transverse profile and (3) the super-Gaussian transverse profile. The deuteron beams have a similar exponential tail temperature of 20 MeV. It can be seen from Fig. A3 (top left, blue curve) that an exponential deuteron distribution at this temperature produces 0.023 neutron/deuteron/sr within 5° cone. While the deuteron beam in case 2 (orange curve) only gives $2 \times 10^{-3}$ neutron/deuteron/sr within 5° cone, using the super-Gaussian profile (case 3) improves this to $5.5 \times 10^{-3}$ neutron/deuteron/sr (red curve). Furthermore, the low energy deuterons with low neutron production cross sections may be filtered by a solid foil of sufficient stopping power, such that the overall neutron production per deuteron is increased. The effect of such a filter is approximated here by applying a 5 MeV low energy threshold to the simulation deuteron spectra. Fig. A3 shows that this can further increase the normalized forward flux to ~0.08, 0.017 neutron/deuteron/sr for case 2 (black dashed curve) and 3 (grey dashed curve) respectively. The corresponding neutron spectra are also compared where a similar trend is observed. Despite that the foil will filter out a large amount of low energy ions, there is minimal impact on the absolute neutron yield and forward flux in the angular and energy distributions, as shown in Fig. A3 (right column). While backwards directed neutrons are not of interest here, it is noted that the total $4\pi$ neutron emissions will be greater than those reported in $2\pi$.

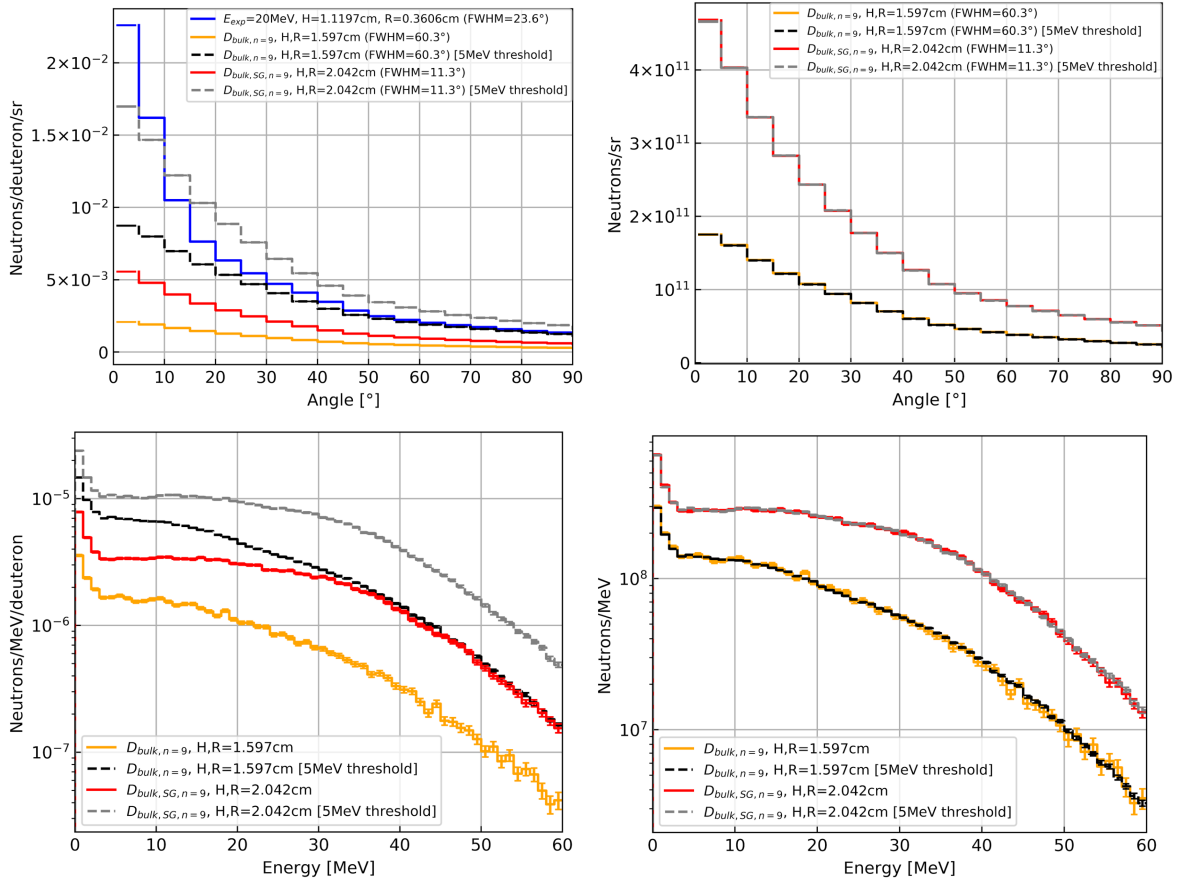

**Figure A3** (*Top*) The angular distributions of the neutron from bulk deuterons in $n = 9$ target. Both normalized (*left*) neutron fluxes and absolute (*right*) yields are shown, as well as for the case a 5MeV energy filter is applied to the deuteron distribution. The 1D ideal normalized flux (blue) is almost recovered by the use of a more uniform laser intensity profile and a filtering foil. (*Bottom*) The corresponding normalized (*left*) and absolute (*right*) neutron spectra show similar effects from the SG laser profile and the filtering foil.